\documentclass[letterpaper,12pt]{article}   
\usepackage{osajnl2} 

\begin{document}

\title{Defocus test and defocus correction in full-field optical coherence tomography}

\author{S. Labiau$^{1,2}$, G. David$^{1,2}$,S. Gigan$^{1,*}$, A.C. Boccara$^1$}
\address{$^1$ Institut Langevin, ESPCI, CNRS UMR 7587, Laboratoire d'Optique Physique, 10 rue Vauquelin, 75231 Paris Cedex 05, France.\\
$^2$ LLTech, Bio-incubateur Eurosant\'e, 70 rue du Dr. Yersin, 59120 LOOS - France\\
$^*$Corresponding author: sylvain.gigan@espci.fr
}

\begin{abstract}
 We report experimental evidence and correction of defocus in full-field OCT of biological samples due to mismatch of the refractive index of biological tissues and water. Via a metric based on the image quality, we demonstrate that we are able to compensate this index-induced defocus and to recover a sharp image in depth.

\end{abstract}

\maketitle 

	Optical Coherence Tomography (OCT) \cite{Fujimoto,Fercher, Bouma} is a powerful technique used to image inside  biological tissues, in particular in the field of eye examination. OCT has been successfully coupled to adaptive optics in order to increase the resolution of retinal examination \cite{Miller,Hermann}. In such experiments, wavefront distortion is due to propagation through a transparent medium (anterior chamber and vitreous humor). The approach proposed here allows us to correct wavefront distortions  induced when light is propagating \textit{within} a highly scattering medium. 
 In contrast with time domain OCT or Fourier domain OCT, Full-Field OCT (FF-OCT) \cite{Dubois,Vabre} directly takes "en face" images. The sample is displaced to position the focal plane at different depths below the surface and to obtain a 3D tomographic image. FF-OCT is in principle able to operate with high lateral resolution using medium or large aperture microscope objectives:  we have used N.A.s ranging from 0.3 to 0.8 with water immersion objectives providing a lateral resolution ranging from about $1.5 \mu m$ to $ 0.5\mu m$ when using a central wavelength of  $750 nm$. 
For time domain, spectral, Fourier-domain or swept-source OCT, the available depth range for imaging is on the order of the depth of field of the optics, therefore requiring low numerical aperture optics that limit the lateral resolution. For example, for a depth of field larger than  $200\mu m$, the N.A. has to be lower than 0.1, giving a lateral resolution larger than  $5\mu m$. Nevertheless in all approaches the axial sectioning ability is linked to the coherence length of the light source.

\begin{figure}[htbp]
  \centering
  \includegraphics[width=3in]{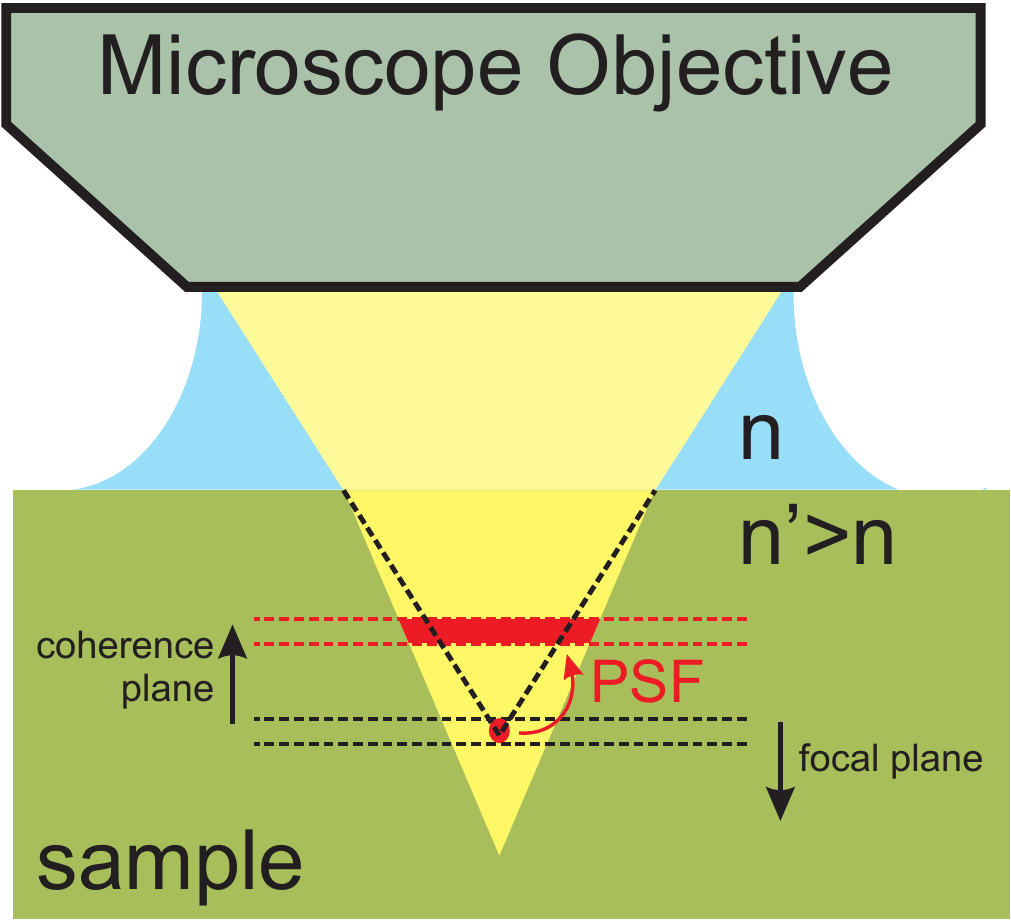}
  \caption{ Principle of the defocusing phenomenon. In a tissue with index $n'$ higher than water, the coherence plane (corresponding to zero path difference) is shifted upwards while the focus is moved downwards.} 
  \label{Defocus}
\end{figure}

In FF-OCT, it is crucial that both arms are as symmetric as possible in term of refractive indices.  As a consequence, FF-OCT generally uses water-immersion microscope objectives, water being the main component of biological tissues. Nevertheless, the refractive index of tissues of medical interest ranges from 1.35 to more than 1.50 \cite{Tearney}. When focusing inside tissues, the focus is shifted forward while the coherence plane goes backward (figure \ref{Defocus}). This phenomenon has been used previously to measure the refractive index of various tissues \cite{Tearney}.
More precisely, in the paraxial approximation, displacing a sample with refractive index $n'$ (instead of $n=1.33$) by a distance $z$ shifts the focal plane forward by
$F(z) = \frac{n'}{n} z$
relative to the sample while the plane of zero path difference is shifted backward (larger refractive index) by 
$
\Pi(z)=-\frac{n}{n'} z
$.
In order to match the position of the focal plane and of the plane of zero path difference  one must introduce a path difference on the reference arm of: 
\begin{equation}
\delta_f(z)=2 n' (F(z)+\Pi(z))=2 n' z (\frac{n'}{n}-\frac{n}{n'})=2 z \frac{n'^2-n^2}{n}
\end{equation}
As long as $\delta_f/2n'$ is smaller than the depth of field there is no need to correct the path difference from the mismatch between the two arms of the Linnik interferometer. With the 0.3 N.A. water immersion objectives used the corresponding depth of field is about 16 micrometers (the depth of field $h$ of a microscope is linked to the aperture angle $u$ via the standard relation $h={1.22 \lambda}/{n \sin u \tan u}$).  

The experimental setup is based on an interference microscope in the Linnik configuration, i.e., a bulk Michelson interferometer with identical microscope objectives in both arms (see \cite{Vabre} for a detailed setup). For this study we used a pair of Olympus, 10x/0.3W, 3.3mm working distance, providing a theoretical transverse (x-y) resolution of $\sim 1.4 \mu m$ at the mean wavelength $\lambda_0= 750 nm$. A halogen lamp is used as a spatially and temporally incoherent light source. A low reflectivity surface ($\sim 2\%$ in water) is placed at the focal plane of the microscope objective of the reference arm in order to maximize the contrast. The object to be imaged is placed in the other arm of the interferometer. The image of the reference mirror and the image of the object are projected with an achromatic doublet lens onto a CCD camera (DALSA 1M15, 15 Hz, 12 bit digitization, 1024x1024 pixels). Due to the source spectrum, the spectral response of the CCD camera and the optical elements of the experimental setup, the useful wavelengths range from approximately 600 nm to 900 nm giving an axial sectioning of less than $1\mu m$. To eliminate all the background  and to keep only the interferometric signal the path difference is modulated by a PZT actuator. 

To correct for defocus around a depth of interest, we have adapted  an approach which has been proposed by Debarre et al. \cite{Debarre} to our en face FF-OCT method. This approach relies on Fourier-Transform image analysis. In essence, the spatial frequency spectrum of an OCT image comprises several regions. The zero frequency region corresponds to the mean image intensity. The highest frequency region corresponds to noises from different origins (speckle, camera noise, photon noise). The intermediary frequency range corresponds mainly to the image  itself, and contains all the informations about the sharp details. When defocus is present, the amount of energy in this intermediary region decreases due to the loss of details induced by the broadening of the point spread function (PSF). Biological samples exhibit multiscale textures and structures observable down to the resolution limit of the optical microscope. In presence of aberrations the spatial frequency spectrum of most samples (that expand from zero spatial frequency to the objective cutoff) is affected by the filtering linked to the PSF degradation. As a metric of the focusing quality, we therefore monitored the total energy in the intermediary region of the Fourier-transformed image. This energy is maximum when the aberration is minimized. Around the optimal position, this metric shows a well-behaved quadratic shape while farther from the optimum position we found it best fitted by a Lorentzian. By adding a known aberration,  we can try  several values of the defocus  and measure the image quality for each of them. This way, we can reconstruct the metric shape (with a minimum of 3 measurements \cite{Debarre}), and infer the optimal value to add to compensate the initial aberration. Note that we only applied this criterion to defocus, but that this image-based optimization method can be used to correct for any kind of aberration, provided it is possible to add a known quantity to a single aberration mode (for instance with a deformable mirror).

\begin{figure}[htbp]
  \centering
  \includegraphics[width=0.9\textwidth]{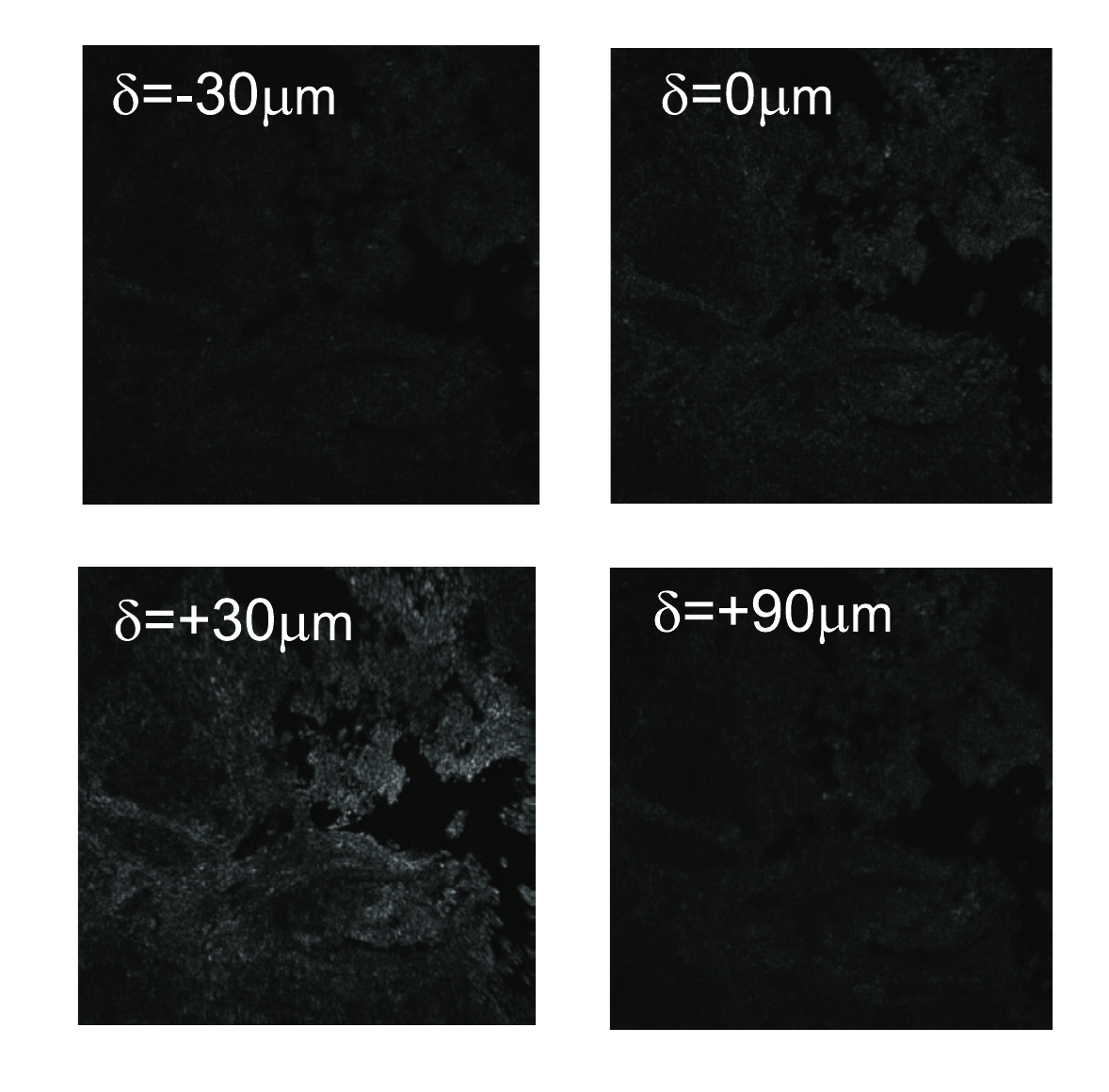}
  \caption{ FF-OCT image  for four different values of added defocus $\delta$. The sample is a sentinel lymph node, the width of the field of view is 500 $\mu m$. The optimal image (for $\delta=+30\mu m$) is both brighter an sharper. } 
  \label{Result1}
\end{figure}

\begin{figure}[htbp]
  \centering
  \includegraphics[width=0.8\textwidth]{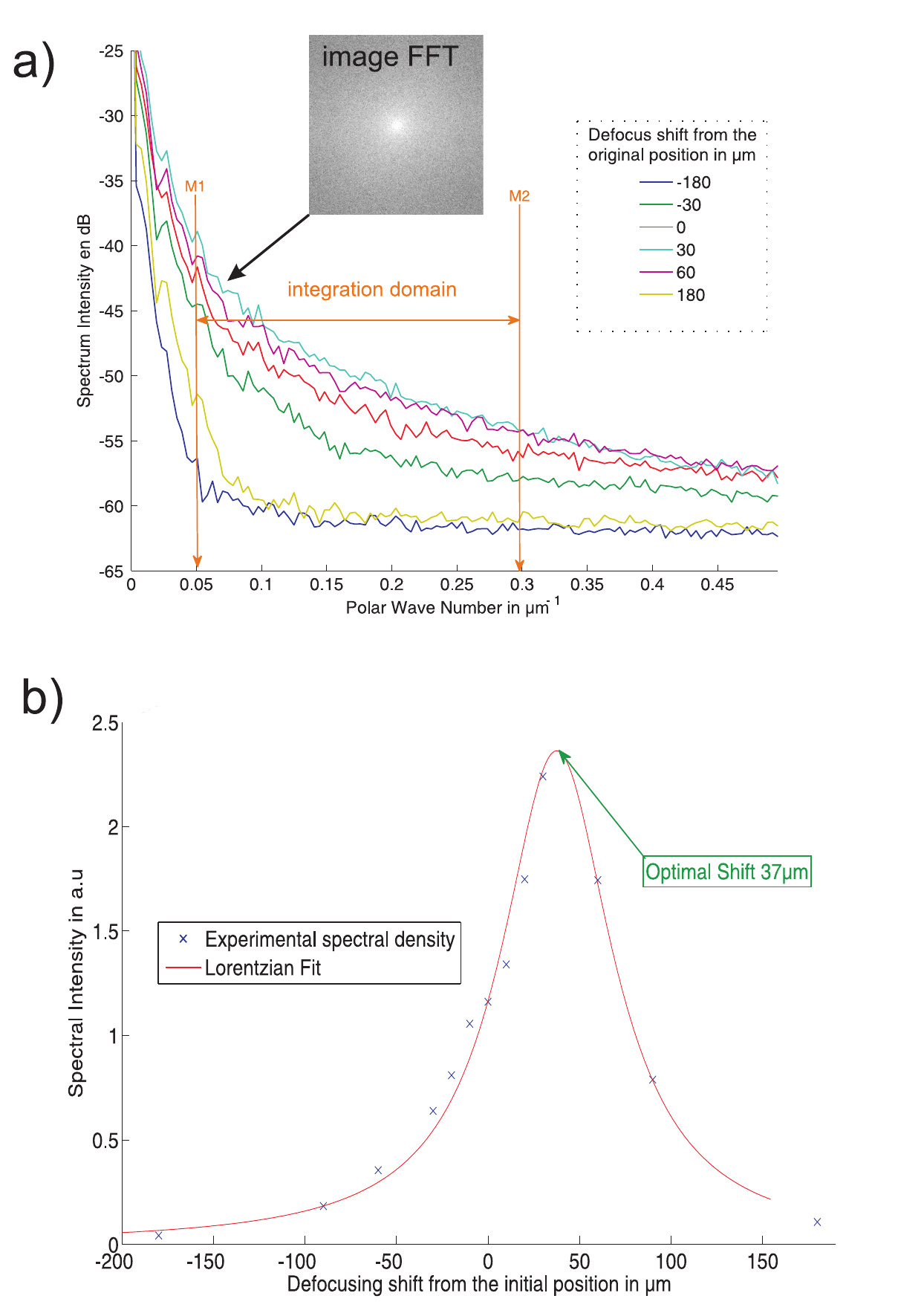}\\
  \caption{ (a) Polar Fourier Spectrum as a function of the amount of defocus introduced. Spectrum are normalized to the mean image intensity. The inset shows the FFT of the optimal image.(b) Figure of merit as a function of defocus $\delta z$ and Lorentzian fit to obtain the optimal focus position.} 
  \label{Result2}
\end{figure}

The experimental procedure was the following. The setup is pre-aligned to ensure focusing of both the reference and the sample arm, and that  the two arms optical paths are matched in water, i.e.,  that the system is correctly aligned for imaging at the surface of a sample.
 We then remove the mirror from the sample arm and move the sample towards the microscope objective, in order to image in-depth. 
  In our experiment, we went up to $200 \mu m $ deep in various tissues samples. Here we describe the results obtained with a sentinel lymph node.  Since the index $n'$ of the sample is larger than the index $n$ of water, defocus starts to degrade the image quality. We  change the length of the reference arm in order to move the coherence plane relatively to the focus of the microscope objective. 
  However, in order to use the metric, it is necessary to always image  the same plane in the sample. Thus,  when changing the reference arm length by $\delta z$ (in air),we need to move the sample in the same direction by a quantity $\delta z/n$, so that the position of the coherence plane in the sample is unchanged. 

For each position of the defocus, we acquired an image, and made a fast-Fourier transform on it. We determine  the upper and lower frequencies of the integration domain to maximise the signal-to-noise ratio. The results obtained are summarized in figures \ref{Result1} and figure \ref{Result2}. $z=0$ corresponds to the   matched arm lengths in water. We find that for our sample, at a depth $z=60\mu m$, we can evaluate from the fit the optimal defocus to be $\delta_z = 37 \mu m$  (well above the depth of field of  $\pm 8 \mu m$). The best image obtained is for $\delta_z \simeq 30 \mu m$, within the focus range, and we see that we are able to recover much more signal, and much sharper details. Note that this allows us to directly estimate in-situ the refractive index of the sample, with a good accuracy. 
Here we find a value of $n=1.52\pm0.01$. The lymph node sample was fixed with formaldehyde, whose index is around $1.42$. The value found seems therefore reasonable, knowing that such a dense tissue \cite{Tearney} originally contains about 55\% of water.

We also see that even though the index mismatch is not large, as the coherence plane moves out-of-focus  linearly with the depth, the image quality is degraded very quickly, even for relatively low numerical aperture. Not taking this effect into account strongly limits the maximum imaging depth achievable in a given sample. OCT images avoid scattering effects which are due to small size  refractive index heterogeneities (less than a few microns),  but larger scale heterogeneities are often present in tissue: they are linked to the geometry of the tissue structure (e.g. fibers) or to the lack of flatness of the sample surface. Moreover the sharp features observed in the corrected image obtained at moderate depth demonstrate that adaptive optics correction is not yet needed at this stage and that defocus was dominant.

For the first time an OCT image from a highly scattering sample has been corrected without wavefront analysis such as in \cite{Denk} but only based on the optimization of the image quality. Compared to the results of \cite{Debarre} we have been able to correct an image captured well below the sample surface of higly scattering media.  This wavefront correction of defocusing has proved to be efficient and simple to implement. 
At this stage we do not know yet if we have reached the optimum  in terms of resolution; from the detailed analysis of cellular structure, it seems that we are close to this goal; nevertheless let us underline that this approach can be extended to other kinds of aberrations when coupled to a deformable mirror conjugated with the pupil of the microscope objective.

We acknowledge the financial support of ANR-TECSAN, ANR MICADO, and Israel-France program "medical and biological imaging". We thank Jonas Binding, Charles Brossollet, Olivier de Witte and Bertrand de Poly for stimulating discussions.

\end{document}